\documentclass[preprint,showpacs,preprintnumbers,amsmath,amssymb]{revtex4}
\usepackage{graphicx}% Include figure files
\usepackage{dcolumn}% Align table columns on decimal point
\usepackage{bm}% bold math

\begin{document}

\title{Simple modeling of self-oscillation in NEMS}% Force line breaks with \\

\author{A. Lazarus}
%\email[]{anthony.ayari@lpmcn.univ-lyon1.fr}
%\homepage[]{Your web page}
%\thanks{}
%\altaffiliation{}
\affiliation{Laboratoire d'Hydrodynamique, \'Ecole Polytechnique,
91128 Palaiseau, France}

\author{T. Barois}
%\homepage[]{Your web page}
%\thanks{}
%\altaffiliation{}
\affiliation{Laboratoire de Physique de la Mati\`ere Condens\'ee
et Nanostructures Universit\'e Lyon 1; CNRS, UMR 5586 Domaine
Scientifique de la Doua F-69622 Villeurbanne cedex, France }
\author{S. Perisanu}
\affiliation{Laboratoire de Physique de la Mati\`ere Condens\'ee
et Nanostructures Universit\'e Lyon 1; CNRS, UMR 5586 Domaine
Scientifique de la Doua F-69622 Villeurbanne cedex, France }
\author{P. Poncharal}
\affiliation{Laboratoire de Physique de la Mati\`ere Condens\'ee
et Nanostructures Universit\'e Lyon 1; CNRS, UMR 5586 Domaine
Scientifique de la Doua F-69622 Villeurbanne cedex, France }

\author{P. Manneville}

\affiliation{Laboratoire d'Hydrodynamique, \'Ecole Polytechnique,
91128 Palaiseau, France}

\author{E. de Langre}
%\email[]{anthony.ayari@lpmcn.univ-lyon1.fr}
%\homepage[]{Your web page}
%\thanks{}
%\altaffiliation{}
\affiliation{Laboratoire d'Hydrodynamique, \'Ecole Polytechnique,
91128 Palaiseau, France}

\author{S. T. Purcell}
\affiliation{Laboratoire de Physique de la Mati\`ere Condens\'ee
et Nanostructures Universit\'e Lyon 1; CNRS, UMR 5586 Domaine
Scientifique de la Doua F-69622 Villeurbanne cedex, France }
\author{P. Vincent}
\author{A. Ayari}
\email[]{anthony.ayari@lpmcn.univ-lyon1.fr}
%\homepage[]{Your web page}
%\thanks{}
%\altaffiliation{}
\affiliation{Laboratoire de Physique de la Mati\`ere Condens\'ee
et Nanostructures Universit\'e Lyon 1; CNRS, UMR 5586 Domaine
Scientifique de la Doua F-69622 Villeurbanne cedex, France }

\date{\today}%

\begin{abstract}

We present here a simple analytical model for self-oscillations in
nano-electro-mechanical systems. We show that a field emission
self-oscillator can be described by a lumped electrical circuit
and that this approach is generalizable to other electromechanical
oscillator devices. The analytical model is supported by dynamical
simulations where the electrostatic parameters are obtained by
finite element computations.

\end{abstract}


\maketitle

Nano-electro-mechanical systems (NEMS)\cite{Craighead2000} are
under extensive research owing to their potential for radio
frequency communication and highly sensitive sensors. This
research, before becoming applicable, will have to cope with
several major issues such as crosstalk. Since the work of
ref.~\onlinecite{Ayari2007}, a new class of NEMS has been
experimentally demonstrated that could circumvent this drawback by
nano-active feedback. In contrast to quartz-oscillator like
architecture,\cite{Colinet2009} there is no need for macroscopic
external active circuit since the nanodevice itself is placed in a
self-oscillating regime. This concept was first theoretically
proposed for NEMS by Gorelik {\it et al.}\cite{Gorelik1998} in the
specific case of the charge shuttle and is now observed in a large
variety of experimental
configurations\cite{Ayari2007,Blick2010,Ionescu2009,Victor2009,VDZ2009,Steeneken2009}.
Although the work of ref.~\onlinecite{Ayari2007} reaches
qualitative agreement between experiment and modelling of the
self-oscillation phenomenon, it lacks simple arguments about the
origin of the instability. Here, we derive a simple linearized
model and an equivalent purely electrical circuit that helps one
getting further insight on the way to design and scale down such
an oscillator. This model is then validated by dynamical and
finite element simulations. The idea exposed in this article, with
minor adaptations, could be useful for other experimental
geometries.

\begin{figure}
\includegraphics[width=7cm]{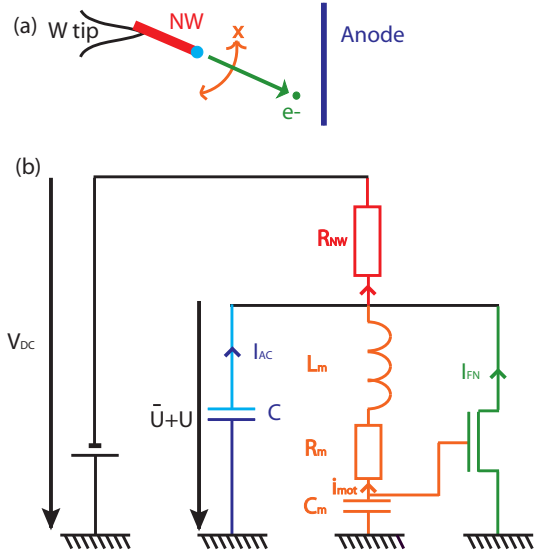}
\caption{(Color online) (a) Schematic of the experimental
configuration and (b) schematic of the equivalent purely
electrical circuit of the self-oscillation of the nano electro
mechanical system of ref.~\onlinecite{Ayari2007}.\label{elec}}
\end{figure}

In a typical experiment, a nanowire (NW) or nanotube with
resistance $R_{\rm NW}$ is attached to a tungsten tip in front of
an anode connected to the ground [Fig.~\ref{elec}(a)]. The tip is
at a negative DC voltage $-V_{\rm DC}$ from the ground; electrons
are emitted from the apex of the nanowire by field emission and
collected by the anode. The NW starts to oscillates spontaneously
in the transverse direction when $V_{\rm DC}$ is larger than some
voltage threshold. This system can be modeled by two coupled
differential equations (see Eq. 1-2 in
ref.~\onlinecite{Ayari2007}): first, a mechanical equation that
can be linearized as follow:
\begin{equation}
\label{Eq1} \ddot{x}+\frac{\omega_0}{Q}\dot{x}+\omega_0^2x=
H\bar{U}U,
\end{equation}
where $x$ is the transverse displacement of the apex of the NW
compared to the equilibrium position (taken positive when the NW
approaches the anode), $2\pi \omega_0$ the resonance frequency of
the mechanical oscillator, $Q$ the quality factor and $H$ a
positive parameter characterizing the actuation strength by
electrostatic forces between the wire and the anode. These
parameters are supposed to be relatively constant in the range of
interest. $\bar{U}$ is the DC voltage between the NW and the anode
and $U$ the AC voltage. $\bar{U}$ is not equal to $V_{\rm DC}$ as
a result of the voltage drop through the nanowire. Second, the
linearized electrical equation reads:
\begin{equation}
\label{Eq2} \left(\frac{\partial I_{FN}}{\partial
U}+\frac{1}{R_{NW}}\right) U+ C\dot{U}=-\frac{\partial
I_{FN}}{\partial x}x-C'\bar{U}\dot{x}\,,
\end{equation}
where $C$ is the capacitance between the NW and the anode, $C'$
its derivative with respect to position, and $I_{\rm
FN}(U+\bar{U};x)$ the field emission current described by the
Fowler--Nordheim equation $I_{\rm
FN}=A(U+\bar{U})^2\beta^2\exp(-B/(U+\bar{U})\beta)$. The $x$
dependence of $I_{\rm FN}$ comes from the field enhancement factor
$\beta$.

An important point to notice is that the field emission
characteristics depends on two inputs, the apex voltage and its
position, in the same way as a transistor or a vacuum tube, but
the role of the gate or grid is played by the spatial degree of
freedom $x$.  A simple equivalent electrical circuit is shown in
Fig.~\ref{elec}(b). The electro-mechanical resonator is
represented by a series RLC circuit in parallel with the capacitor
$C$ of Eq.~\ref{Eq2}. In this well-known analogy, the motional
current through the RLC circuit is $i_{\rm mot}=C'\bar{U}\dot{x}$
and the passive components are the motional inductance $L_{\rm
m}=1/(H\bar{U}^2C')$, the motional resistance $R_{\rm
m}=\omega_0/(QH\bar{U}^2C')$ and the motional capacitance $C_{\rm
m}=H\bar{U}^2C'/\omega_0^2$. The voltage across the motional
capacitance is proportional to $x$ and can be used as the gate
voltage of an equivalent transistor delivering the same field
emission current for a given $x$ and $U+\bar{U}$. The
transconductance of such transistor is $(\partial I_{\rm
FN}/\partial x) H\bar{U}/\omega_0^2$. It brings the gain necessary
to sustain the self-oscillation regime and acts as a feedback
loop.

The main parameter of the self-oscillating circuit is the driving
DC voltage above which the system spontaneously generates the AC
signal. In the following, we derive a simple analytical formula
giving the self-oscillation condition. If the nanowire resistance
$R_{\rm NW}$ is smaller than the field emission resistance
$(\partial I_{\rm FN}/\partial U)^{-1}$, to first order the
voltage at the apex $\bar{U}$ is $V_{\rm DC}$ and there is no
self-oscillation.  We consider the opposite case $R_{\rm
NW}\gg(\partial I_{\rm FN}/\partial U)^{-1}$ because it gives a
simpler formula (the general case can be calculated
straightforwardly by the same method). However, when the nanowire
resistance gets larger more power is dissipated in heating instead
of sustaining the oscillation, so that it might seem optimal to
keep $R_{\rm NW}$ larger than the field emission resistance by
less than an order of magnitude. A single differential equation of
the full electro-mechanical system can be obtained by combining
Eqs.~\ref{Eq1} and~\ref{Eq2}:
\begin{eqnarray}
&&\tau\dddot{x}+\ddot{x}\left(1+\frac{\omega_0\tau}{Q}\right)+
\dot{x}\left(\frac{\omega_0}{Q}+ H\bar{U}^2\tau\frac{\partial\ln
C}{\partial x}+
\omega_0^2\tau\right)\nonumber\\
&&\qquad\qquad\qquad\qquad +x\left(\omega_0^2+H\bar{U}^2
\frac{\partial \ln\beta}{\partial x}\right)=0
\end{eqnarray}
where $\tau=C(\partial I_{\rm FN}/\partial U)^{-1}$ is the
discharge time constant of the electrical circuit. According to
the Routh--Hurwitz criterion this dynamical system is stable when:
\begin{eqnarray}
&&H\bar{U}^2\tau\left[\frac{\partial\ln\beta}{\partial x}
-\frac{\partial\ln C}{\partial x}\left(1+\frac{\omega_0\tau}{Q}\right)\right]\nonumber\\
&&\qquad\qquad\qquad-\frac{\omega_0\tau}{Q}\left[\frac{1}{\tau}+\tau\omega_0^2+\frac{\omega_0}{Q}\right]\geq0
\end{eqnarray}
From this inequality, since $C$ and $\beta$ increase with $x$,
only the variation of $\beta$ with $x$ favors the self-oscillation
regime and we can distinguish between two categories of terms that
prevent from reaching it: {\it i\/}) the variation of the
capacitance with $x$ and {\it ii\/}) the relative value of $\tau$
and $\omega_0^{-1}$. The latter can be minimized for
$\omega_0\tau\sim1$ as long as $Q\gg1$ (our nanowire
resonators\cite{Sorin2007} routinely reach $Q>10^5$). In these
conditions, the geometry of the device should be such that
$\partial\ln(\beta/C)/\partial x>0$ to have a chance to observe
self-oscillations. Finally the threshold DC voltage at the apex
for self-oscillation is:
\begin{equation}
\label{Eqseuil}
\bar{U}_{so}=\frac{\omega_0}{\sqrt{QH\partial\ln(\beta/C)/\partial
x}}
\end{equation}
and the threshold DC voltage of the power supply is $V^{\rm
DC}_{\rm so}=\bar{U}_{\rm so}+ R_{\rm NW}I_{\rm FN}(\bar{U}_{\rm
so},\beta)$.

In order to check the different hypotheses made, we performed
numerical simulations and determined the electrostatic force,
capacitance and field enhancement factor by finite element methods
(FEM). The sample is a straight 10~$\mu$m-long nanowire of radius
100~nm attached to a metallic conical tip in front of a metallic
plate perpendicular to the axis of the tip. The nanowire is
initially tilted by $20^\circ$ compared to the cone axis. The sole
degree of freedom of the nanowire is this angle that can decrease
due the attractive electrostatic force between the wire and the
metallic plate. The distance between the tip end and the plate is
60~$\mu$m. The mechanical restoring force is taken from the
calculated rigidity of a clamped free beam with a Young modulus of
400~GPa and density of 3200~kg/m$^3$, $Q=10^4$ and $R_{\rm
NW}=10^{10}\Omega$. Further details about the simulations and a
more refined mechanical model can be found in
ref.~\onlinecite{Lazarus2010}.
\begin{figure}
\includegraphics[width=7cm]{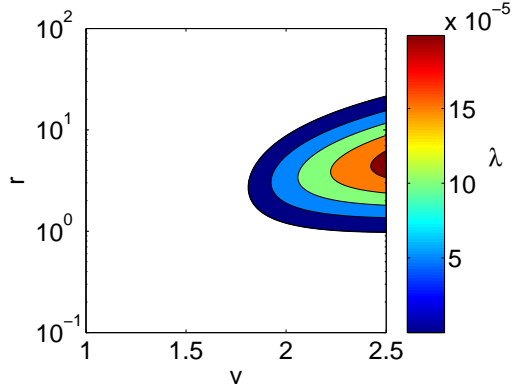}%
\caption{(Color online) Stability map of a nanowire during field
emission for $Q=10000$ and different normalized voltages $v$ and
dimensionless intrinsic
frequencies $r$.\label{map1}}%
\end{figure}
We first simulated the spatial variation of $C$ and $\beta$ and
verified that $\partial\ln(\beta/C)/\partial x>0$ for a wide range
of angles around $20^\circ$, and established that $H$ is changing
by less than 15\%. The dimensionless differential equations were
then rewritten, their eigenvalues computed, and the sign of their
real part $\lambda$ scrutinized. The real part defines the growth
rate of the mode and the solution, which is proportional to
$\exp(\lambda t)$, decay to zero when it is negative, so that the
system is stable. On the contrary, $\lambda>0$ makes the system
unstable and leads it into a stable self-oscillating regime thanks
to nonlinear saturating terms. The oscillation amplitude gets
larger as $\lambda$ increases. Finally, we determined stability
maps giving the parameter regions where  $\lambda$ is positive and
self-oscillations possible.

Fig. \ref{map1} represents the stability map of the system for
different applied DC voltages $v=V_{\rm DC}/V_{\rm ref}$ and
different dimensionless intrinsic frequencies $r=\omega_0\tau$.
$V_{\rm ref}=400\,$V is the voltage above which $R_{\rm NW}$ stops
being negligible when compared to the field emission resistance.
One can point out that {\it i\/}) there is no self-oscillation for
$v\ll1$, {\it ii\/})  self-oscillations are easier at higher $v$
(the growth rate is larger and the instability region wider). This
validates the statement that for optimal self-oscillations $R_{\rm
NW}$ needs to be bigger than the field emission resistance (the
field emission current increases exponentially with $v$, so that
the field emission resistance is smaller for higher $v$). This
figure also clearly demonstrates that self-oscillations are
obtained at easiest for $r\sim1$.

\begin{figure}
\includegraphics[width=7cm]{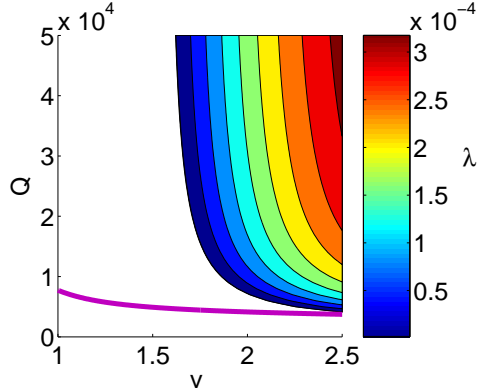}%
 \caption{(Color online) Stability map of a nanowire during
field emission for a dimensionless frequency $r=5$ and different
normalized voltages $v$ and $Q$. The solid line represents the
self-oscillation threshold determined using
Eq.~\ref{Eqseuil}.\label{map2}}
\end{figure}
We also calculated the stability map for various quality factors.
Eq.~\ref{Eqseuil} that determines the boundary between the stable
region and the self-oscillation region is in relatively good
agreement with the results of numerical simulations for high
voltage, i.e. when $\partial I_{\rm FN}/\partial U\gg1/R_{\rm
NW}$. This confirms the validity of the above analytical
derivation. Eq.~\ref{Eqseuil} shows that, as for any other NEMS
device, keeping good performance (in this case by maintaining the
operating voltage low) at the nanoscale and high frequency
requires an improvement of the capacitive coupling and the quality
factor. Finally a simple scaling calculation shows that $r$
decreases like the inverse of the apex-anode distance. Downscaling
thus helps one to reach the regime where $r\sim1$. If this term
become too small, or if the resistance of the nanowire or nanotube
saturates in the ballistic regime, the device can still be
operated with the help of an additional constant resistance
between the DC power supply and R$_{\rm NW}$.

In conclusion, using an electrical equivalent circuit, we showed
that the origin of self-oscillations in field emission NEMS can be
understood in terms of motional capacitance and spatial variation
of the field emission current in a feedback loop. An equation  was
derived to determine the threshold voltage for self-oscillation
and its output confirmed by numerical and FEM simulations. We
expect that our simple model will demystify the mechanism
responsible for self-oscillation in field emission NEMS, as it
appears that it can be understood with simple classical electrical
passive components and one transistor. It appears then that
geometries like the one of ref.~\onlinecite{Ionescu2009} where the
self-oscillation mechanism is not yet clearly identified are
indeed very similar to ours and may be understood within the same
framework.  This work opens up perspectives for the control and
fabrication of low power nano-oscillators for time base and AC
generators applications.

\begin{acknowledgments}
This work was supported by French National Research Agency
(NEXTNEMS : ANR-07-NANO-008-01 and AUTONOME : ANR-07-JCJC-0145-01)
and R\'egion Rh\^one-Alpes {\sl CIBLE\/} program. The authors
acknowledge the ``plateforme nanofils et nanotubes lyonnaise''.
\end{acknowledgments}

% Create the reference section using BibTeX:

%Merlin.mbs v4.21 2009-07-09.
%\bibliography{autoAPL3}

\end{document}